\documentstyle[prl,aps,epsf]{revtex}

\begin{document}

\twocolumn[\hsize\textwidth\columnwidth\hsize\csname@twocolumnfalse\endcsname
 
\title{Modification of the Landau-Lifshitz Equation in
  the Presence of a Spin-Polarized Current in CMR and GMR Materials. }

\author{Ya.B.Bazaliy, B.A.Jones, and Shou-Cheng Zhang}

\address{Department of Physics, Stanford University, Stanford, CA}

\address{IBM Almaden Research Center, San Jose, CA}

\date{\today}
\maketitle
\begin{abstract}
  We derive a continuum equation for the magnetization of a conducting
ferromagnet in the presence of a spin-polarized current. Current
effects enter in the form of a topological term in the Landau-Lifshitz
equation . In the stationary situation the problem maps onto the
motion of a classical charged particle in the field of a magnetic
monopole. The spatial dependence of the magnetization is calculated
for a one-dimensional geometry and suggestions for experimental
observation are made. We also consider time-dependent solutions and
predict a spin-wave instability for large currents.
\end{abstract}

]

\tighten

Phenomena associated with spin-polarized currents in layered materials
and in Mn-oxides have attracted high interest recently. Efforts are
strongly concentrated on theoretical and experimental investigation of
large magnetoresistance, which is of great value for future
applications. Examples of the effect are GMR in layered materials (see
review \cite{GMR}), spin valve effect for a particular case of a
three-layer sandwich and CMR in the manganese oxides (see review
\cite{CMR}). 

The dependence of resistivity on magnetic field is explained
conceptually in two steps: first the magnetic field changes the
magnetic configuration of the material and that in turn influences the
current.  Of course, as for any interaction there must be a
back-action of the current on the magnetic structure. The existence of
such back-action was explored in \cite{slon,berger1}. Several
current-controlled micro-devices utilizing this principle were
proposed \cite{slon}.  In both papers layered structures with
magnetization being constant throughout the magnetic layers were
considered. In the present paper we derive the equations for a
continuously changing magnetization in the presence of a
spin-polarized current. This equation takes the form of a
Landau-Lifshitz equation with an additional topological term, and
admits a useful analogy with a mechanical system. We discuss several
solutions in one-dimensional geometries. Our equations also can be
viewed as a continuum generalization of \cite{slon,berger1} for layer
thickness going to zero.

Consider a current propagating through a conducting ferromagnet.
Conducting electrons are viewed as free electrons interacting only
with local magnetization $\bf M$. The motion of each individual
electron is governed by the Schroedinger equation with a term $J_H
\mbox{\boldmath $\sigma$}{\bf M}$, where $J_H$ is the value of the
Hund's rule coupling or in general of the local exchange.  Since
spin-up electrons have lower energy a nonzero average spin of
conducting electrons $(1/2)\langle\mbox{\boldmath$\sigma$}\rangle $
develops. An angular momentum density $(\hbar/2)
\langle\mbox{\boldmath$\sigma$}\rangle$ is then carried with the
electron current so we have a flux of angular momentum.  This leads to
a non-zero average torque acting on the magnetization which can
deflect it from the original direction (see figure~1).

\begin{figure}[hbt]
\epsfxsize=5cm\epsfbox{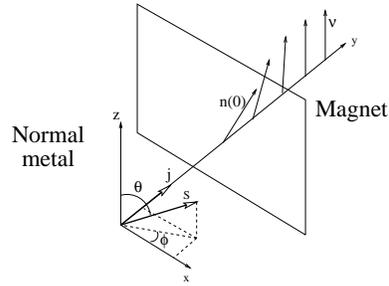}
\vspace{15pt}
\caption{Experimental setting: spin-polarized current enters a 
  half-infinite magnet from the left. Originally the magnetization is
  aligned along the easy-axis $\nu\| z$. However
  if the incoming electrons are spin-polarized in a different
  direction, their interaction with the magnetization leads to a
  deflection of the magnetization.}
\end{figure}

Propagation of a current in a ferromagnet should be described by a
system of two equations: one for the motion of conducting electrons
and another for the magnetization. We derive here the second equation
in the limit of small space-time gradients and present several
solutions. The case of very large $J_H\to \infty $ is considered,
meaning a complete polarization of electron spins in the ferromagnet.
This case can be often realized  in experiment. In the layered
structures magnetic layers can be made of a material with large
band splitting, like Heusler alloys, in which the spin opposite to
magnetization direction can not propagate. In the CMR materials large
Hund's rule coupling is well known \cite{CMR} and constitutes the basis for a
double-exchange mechanism governing their magnetic ordering.


{\bf Schroedinger Equation}: the conducting electrons are considered
noninteracting with an $\varepsilon=p^2/2m$ energy spectrum:

\begin{equation}
\label{schrodinger-eq}i\hbar \frac{\partial \psi _\alpha }{\partial t}=-%
\frac{\hbar ^2}{2m}\Delta \psi _\alpha +
J_H {\bf M}({\bf r},t) \mbox{\boldmath$\sigma$}
_{\alpha \beta }\psi _\beta  
\end{equation}
We diagonalize the matrix
${\bf~M(r},t)\mbox{\boldmath$\sigma$}_{\alpha \beta }$ with a local
spin rotation $\phi_{\alpha} =
{U}_{\alpha\beta}({\bf~r},t)\psi_{\beta}$.  The spinor $\phi$
describes the electron in the coordinate system with $z$-axis being
parallel to the local magnetization.  Retaining only the first order
terms in gradients and using $\phi_{-}\to 0$ for $J_H\to \infty$ we
reduce (\ref{schrodinger-eq}) from a system of two equations to
one equation for $\phi_{+} \equiv \phi$ spin amplitude:

\begin{equation}
\label{phi-equation}i\hbar \frac{\partial \phi}{\partial t}=-\frac{
\hbar ^2}{2m}\Delta \phi +J_HM\phi -\frac{\hbar ^2}{2m}(U_{+\beta }\nabla
_iU_{\beta +})\nabla _i\phi 
\end{equation}
The last term in (\ref{phi-equation}) can be transformed (${\bf M} = M
{\bf n}$ ):

\begin{equation}
\label{schrodinger-with-monopole-A}i\hbar \frac{\partial \phi }{\partial t}
=- \frac{\hbar ^2}{2m}\Delta \phi +J_H M\phi -\frac{i\hbar ^2}{4m}
A_{\rm mon}^k({\bf n}) \nabla _in^k\nabla _i\phi 
\end{equation}
where ${\bf A}^{\rm mon}({\bf n})$ is a function satisfying the
following equations:

$$
\epsilon _{\alpha \beta \gamma }\frac{\partial A_\beta ^{\rm mon}}{\partial
n_\gamma }=2n_\alpha \ ,\;\;\frac{\partial A_\alpha ^{\rm mon}}{\partial n_\beta 
}=\epsilon _{\alpha \beta \gamma }n_\gamma  
$$
If we view $n^2({\bf r},t)=1$ as a sphere, ${\bf A}^{\rm mon}$ has
the simple interpretation of the vector potential due to a magnetic
monopole located at the center of the sphere.  The monopole term is
known to appear from $U_{+\beta }\nabla_i U_{\beta +} $ in the theory
of Berry phase and is used by other CMR theories in different forms
(see \cite{berry}).  Equation (\ref{schrodinger-with-monopole-A}) has
a form of a Schroedinger equation in a magnetic field expanded up to
the linear term in ${\bf A}^{\rm eff}$, with vector potential
$A_i^{\rm eff}=(i\hbar ^2/4m)A_{\rm mon}^k\nabla _in^k$. It describes
the motion of the conducting electrons in the given field ${\bf
  n}({\bf r},t)$.  Conversely it gives the interaction between the
current and the magnetization. The form of equation is the same as for
an electromagnetic interaction, and hence we can write by analogy:
\begin{equation}
\label{interaction-term}
E_{\rm int}=\frac 1cj_iA_i^{\rm eff}=
    -\frac \hbar 4\frac{j_i}eA_{\rm mon}^k({\bf n})\nabla _in^k
\end{equation}
where ${\bf j}$ is an electric current.


{\bf Magnetization motion} is described by  Landau-Lifshitz
equations which are obtained from the energy functional.  After adding
(\ref{interaction-term}) to the usual energy density of a ferromagnet
with uniaxial anisotropy along the axis \mbox{\boldmath{$\nu$}}, we obtain:
\begin{equation}
\label{energy-functional}
E=\int [J({\bf \nabla M})^2 - K(\mbox{\boldmath{$\nu$}}
{\bf n})^2+\frac 1cj_iA_i^{eff}] dV
\end{equation}
with $K>0$ corresponding to easy-axis and $K<0$ to easy-plane magnets.
The equations of motion then take the form:
\begin{equation}
\label{l-l}
\frac{\partial {\bf M}}{\partial t}= 
    \frac{ g \mid e \mid}{ 2 m c} [{\bf f}\times {\bf M}] ] 
\end{equation}
\begin{equation}
\label{n-equation-final}
{\bf f}  = -\frac{\delta E}{\delta {\bf M}} =
     J\Delta {\bf M} + \frac{2K}{M^2}({\bf \mbox{\boldmath{$\nu$}} M})\mbox{\boldmath{$\nu$}} +
     \frac{\hbar}{2 M} \frac{j_i}{e} [\nabla_i {\bf n} \times {\bf n}]
\end{equation}
where the last term in $\bf f$ is new and describes the effect of the
current.  The system of (\ref {schrodinger-with-monopole-A}) and
(\ref{l-l},\ref{n-equation-final}) constitute a complete set of
equations for a magnet with current. Equations
(\ref{l-l},\ref{n-equation-final}), generalizing the Landau-Lifshitz
equation in the presence of a current, are the central result of this
work.

Since magnetization corresponds to  angular momentum ${\bf L}=\hbar
/g\mu _B\;{\bf M}$, an equation of the angular momentum flux
continuity follows from (\ref{n-equation-final}):

\begin{equation}
\label{}
\frac{\partial L_k}{\partial t}+\frac{\partial \Lambda _{ki}}{\partial x_i}
=\left[\frac{2 g\mu_B K}{\hbar M^2}({\bf \mbox{\boldmath{$\nu$}}
      n}) \mbox{\boldmath{$\nu$}} \times {\bf M}\right] _k 
\end{equation}

\begin{equation}
\label{L-equation-flux-form-B}\Lambda _{ki}=M^2J[{\bf n}\times \nabla _i{\bf %
n}]_k+\frac{\hbar}{ 2}(\frac{j_i}{e} n_k) 
\end{equation}
The flux $\Lambda_{ki}$ consists of two parts: one due to the spatial
derivatives of magnetization and another due to the motion of
conducting electrons. In our situation the spins of moving electrons
are parallel to ${\bf n}$. That is why their contribution is
factorized in the form $(\hbar /2)(j_i/e)n_k$.


{\bf Consider the stationary case} in an experimental setting shown on
figure~1.  For the stationary process the r.h.s. of equation
(\ref{l-l}) vanishes. The current propagates along the ${\hat{\bf y}
  }$ direction.  All spatial derivatives reduce to
$\mbox{\boldmath$\nabla$} \rightarrow \nabla _y $. For the reasons
immediately following we will denote differentiation with a prime to
get a resemblance to a time derivative in notation $\nabla_y
{\bf~n}\equiv {\bf n'}$. From (\ref{n-equation-final}) we get an
equation on ${\bf n}({\bf r},t)$:

\begin{equation}
\label{stationary-const-j-equation-modified}{\tilde J}[{\bf n}''\times 
{\bf n}]=[(-{\tilde K}({\bf  \mbox{\boldmath{$\nu$}}   n}) \mbox{\boldmath{$\nu$}}  -Q(\frac je)[{\bf n'
}\times {\bf n}])\times {\bf n}] 
\end{equation}
with new parameters:
$$
\tilde J=\frac{g\mu _BM}\hbar J,\;\;\tilde K=\frac{2g\mu _B}{\hbar M}K,\;\;Q=\frac{g\mu _B}{2M} 
$$
Since $\bf n$ in the stationary case depends on $y$ only, we can
interpret $y$ as a fictitious time; together with ${\bf n}^2=1$
equation (\ref{stationary-const-j-equation-modified}) can then be
interpreted as the equation of motion for a particle of a mass $\tilde
J$ confined to the surface of a unit sphere and experiencing two
forces: \\
(a) a force of magnitude $-{\tilde K}({\bf \mbox{\boldmath{$\nu$}} n})
\mbox{\boldmath{$\nu$}}$ parallel to the
anisotropy axis, \\
(b) a Lorentz force , due to a field ${\bf H}_{\rm mon}=-Q(\frac
je){\bf n}$ of a magnetic monopole positioned in the
center of the sphere.\\
The vector product ensures that only tangential components of the
total force act on the particle. The normal component is compensated
by the reaction forces. Such an anology enables one to visualize the
soluctions of the original equation (\ref
{stationary-const-j-equation-modified}) as trajectories of a massive
particle on the sphere.

The equation of particle motion in the field of a magnetic monopole (\ref
{stationary-const-j-equation-modified}) has two first integrals \cite
{particle-in-monopole-field}. 
\begin{equation}
\label{energy-integral}W=\frac{\tilde J{\bf n'}^2}2+\frac{{\tilde K}( 
{\bf  \mbox{\boldmath{$\nu$}}   n})^2}2={\rm const} 
\end{equation}
\begin{equation}
\label{ang-momentum-integral}D_\nu ={\tilde J}([{\bf n}\times {\bf n'}
]\nu )+Q(\frac je)({\bf n \mbox{\boldmath{$\nu$}} })={\rm const} 
\end{equation}
Together they give a way to solve
(\ref{stationary-const-j-equation-modified} ) for arbitrary initial
conditions.  Expressing everything through the Euler angles $\{\phi
(y),\theta (y)\}$ (defined on figure 1) of the vector $\bf n$ , we
obtain: 
$$
\phi' =\frac{D_\nu -Q(j/e)\cos \theta }{\tilde J\sin ^2\theta } 
$$

\begin{eqnarray}
\label{first-integrals-used}
\nonumber
\theta' & = & \sqrt{ 
   \frac{2W-\tilde K \cos ^2\theta}{\tilde J} 
  -\frac{(D_\nu -Q(j/e)\cos \theta )^2}{{\tilde J}^2\sin ^2\theta }
 }\\
  & \equiv & {\cal F}(\theta ) 
\end{eqnarray}
The problem for $\theta $ is solved by the implicit function: 
\begin{equation}
\label{theta-solution} y = y(0)+\int_{\theta (0)}^\theta \frac{d\theta }{
{\cal F}(\theta )} 
\end{equation}
afterwards $\phi $ can be found from the first equation in (\ref
{first-integrals-used}).

Assume that deep inside the magnet ($y \to \infty $) the magnetization
resumes its original direction along the anisotropy axis ${\bf n}\to
\mbox{\boldmath{$\nu$}}, \;\; {\bf n}'\to 0$.
From this the values of the first integrals can be found and
substituted into (\ref{first-integrals-used}).  Natural length and
current scales appear in the calculation:

\begin{equation}
L_m =\sqrt{\frac{JM^2}{8K}},\;\;\;j_0=\frac{e}{\hbar}
\sqrt{KJM^2}
\end{equation}
through which the material parameters $J,K,M$ enter the problem.
Their values for different materials are given in the following table:

\vspace{2mm}

\begin{tabular}{|c|c|c|c|}
\hline
Material & \mbox{$L_m \;\; [\AA]$} & \mbox{$j_0\;\;[{\rm A/cm^2}$}] & Ref. \\
\hline
\mbox{CMR: \mbox{$\rm La_{0.66}Ca_{0.33} MnO_3$}}  & \mbox{$> 130$}  &
\mbox{$< 4 \cdot 10^7$} & \protect\cite{lynn} \\
\hline
Fe  &  40  & \mbox{$1.1 \cdot 10^8$} & \protect\cite{jiles}\\
\hline
\mbox{Heusler Alloy: PtMnSb}  & \mbox{$50\sim 100$} & \mbox{$\sim 5 \cdot
  10^7$} & \protect\cite{heusler}\\
\hline
\end{tabular}

\vspace{2mm}

The integral (\ref{theta-solution}) can be then expressed in
elementary functions but the formula is long and will be detailed in a
later paper.  Instead, the results are presented on figure~2.  It is
seen that magnetization relaxes in a distance $\approx 10 L_{m}$ which
is about the width of the domain wall in the material.

In the ``particle picture'' the motion starts at some point $A$ on the
trajectory, yet to be determined from the b.c. on the normal metal -
magnet interface, and ends on the North pole. The particle has
just enough energy to climb the potential hill and come to rest on the
top. The particle trajectory is bent by the monopole field.  In the
absence of the monopole the particle would go along the meridian.


{\bf The boundary condition on the metal-magnet interface},
$y=0$ is derived from the continuity of the angular momentum flux.
Such condition ensures that there is no torque concentrated on the
boundary consistent with the assumption of slow spatial changes of
the magnetization.

\begin{figure}[b]
\epsfxsize=8cm\epsfbox{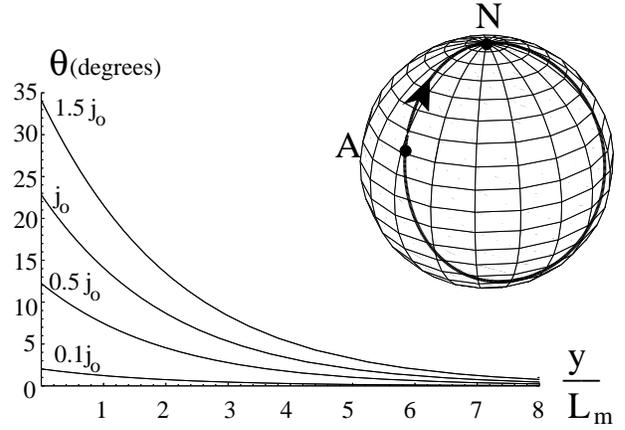}
\vspace{15pt}
\caption{Magnetization deflection angle $\theta(y)$. 
  Curves correspond to different values of current, $\theta_0 =
  \pi/2,\;p=0.5$.  Inset: trajectory of particle on the sphere for
  $j=j_0$. Starting point A corresponds to the conditions of the main
  graph.}
\end{figure}

The reflection of the down-spin electron component occurs on the
length scale of the electron wavelength. On this distance
magnetization is almost constant and solving the one-particle
reflection problem we find the jump of the electron flux component in
the $\hat y$ direction $\Sigma_i \equiv \Sigma_{iy}$ to be:

\begin{equation}
{\bf\Sigma}_{\rm metal}^{\rm electron} - {\bf\Sigma}_{\rm magnet}^{\rm electron} =
{\bf \Sigma }-{\bf n}( 
{\bf \Sigma }{\bf n}) 
\end{equation}
where $\bf\Sigma$ is the average injected flux. From
(\ref{L-equation-flux-form-B}) the flux inside the magnet is:
\begin{equation}
\Lambda_{i,y} = {\bf\Sigma}_{\rm magnet}^{\rm electron} +
                 M^2 J [{\bf n}_0\times {\bf n}'_0]_i 
\end{equation}
In the metal, only the electron part of the flux is present.  Then
continuity gives the boundary condition:
\begin{equation}
\label{boundary-condition}
{\bf \Sigma }-({\bf \Sigma n}_0){\bf n}_0 = M^2 J[{\bf n}_0\times {\bf n}'_0] 
\end{equation}
Note that it involves both the vector ${\bf n}$ and its derivative on the
boundary.

Condition (\ref{boundary-condition}) can be transformed into a
system of two algebraic equations and an inequality:
\begin{eqnarray}\label{algebraic-system}
\nonumber
& &\frac 12\left( \frac{pj}{j_0}\right) ^2(1-y^2) = 1-x^2 \nonumber\\
& &p(\cos \theta _0-xy)=1-x  \\
& & y^2+x^2-2yx\cos \theta _0\leq \sin ^2\theta _0\nonumber
\end{eqnarray}
where ${\bf\Sigma}=\Sigma~{\bf e}$, $\cos\theta_0 = {\bf e \cdot
  \mbox{\boldmath{$\nu$}}}, \; x =
{\bf\mbox{\boldmath{$\nu$}}\cdot~n_0}, \; y = {\bf~e~\cdot~n_0}$.
The parameter $p=e\Sigma/j, \;\; p\in~[0,1/2]$, describes the ``degree
of polarization'' of the incident electrons .  The inequality in
(\ref{algebraic-system}) is a geometrical constraint on $x$ and $y$
arising from their definition.

The trajectory is determined by three parameters:
$(j/j_0,p,\theta_0)$. We can plot a domain of existence of a solution
to (\ref {algebraic-system}) in the 3-D space of these parameters.  A
typical 2-D section of this diagram for constant $\theta _0$ is
shown in figure~3. A solution is absent in the regions B and C which
means that for larger currents and spin-polarizations no smooth
stationary solution approaching the easy-axis direction at infinity is
available.  Either a non-stationary solution or a solution which never
approaches
\mbox{\boldmath{$\nu$}} will be realized in that region.

\begin{figure}[t]
\epsfxsize=5cm\epsfbox{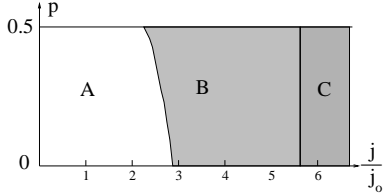}
\vspace{15pt}
\caption{ A typical 2-D section of the phase diagram plotted for
  $\theta_0 = \pi/3$. A: domain of existence of a solution to
  (\protect\ref {algebraic-system}); B: no solution, b.c. at $y=0$ can
  not be satisfied; C: no solution, b.c. at $y\to\infty$ can not be
  satisfied and spin wave instability occurs.}
\end{figure}


{\bf Time-dependent solutions} of equation (\ref {n-equation-final})
can be found in some cases. We again assume the current ${\bf j}$ to
be uniform. We rewrite equation (\ref{n-equation-final}) through 
${\bf n}({\bf r},t)$:

$$
\frac{\partial {\bf n}}{\partial t}=[{\bf g}\times {\bf n}]-Q(\frac{j_i} e%
)\nabla _i{\bf n} 
$$
\begin{equation}
\label{n-equation-newform}{\bf g} = {\tilde J}\Delta {\bf n} + {\ 
\tilde K}({\bf \mbox{\boldmath{$\nu$}}   n})\mbox{\boldmath{$\nu$}} 
\end{equation}
and suppose $n_0({\bf r})$ solves $[{\bf g}\times {\bf n_0}]=0$, i.e.
represents a static solution in the absence of the current.  Then
\begin{equation}
\label{moving-solution}n({\bf r},t)=n_0({\bf r}+Q(\frac{{\bf j}}e)t)=n_0( 
{\bf r}+\frac{\omega _0 L_m }{\sqrt{2}}\frac j{\ j_0}\ t) 
\end{equation}
where $\omega _0 = \tilde K$, is a solution of
(\ref{n-equation-newform}) for a nonzero current. For instance a
moving Bloch wall will be a solution when current is flowing
perpendicular to it (provided pinning is absent).

Another particular solution is a spin wave in the presence of a current.
We search for a solution (\ref{n-equation-final}) in the form of a spin
wave: $\{\theta ={\rm const},\;\phi ={\bf kr}-\omega t\}$.  This gives the
spectrum:
\begin{equation}
\label{spectrum}\omega =\tilde Jk^2-Q\frac{{\bf jk}}e+\tilde K 
\end{equation}
As we see, the current changes the energy gap of spin waves and shifts the
position of the  minimum: 
$$
\omega _{\min }=\tilde K-(Q\frac je)^2\frac{\cos ^2\alpha }{4\tilde J}
=\omega _0(1-\frac{j^2}{32j_0^2} \cos^2 \alpha ) 
$$
where $\alpha$ is the angle between $j$ and $k$ and\ $\omega
_0=\tilde K$ \ is the gap of spin wave in an anisotropic ferromagnet.
For large enough current $j>4\sqrt{2}\ j_0$ an instability occurs.
That is also the condition which leads in the region $C$ on figure~3
to the loss of any trajectory approaching \mbox{\boldmath{$\nu$}} at
infinity as the integral (\ref{theta-solution}) becomes undetermined.
A spin-wave instability is also predicted in other models of
spin-polarized transport \cite{berger1}.


{\bf Discussing possible experiments} we note that the characteristic
current is large, but such densities are in fact common for layered
metallic structures and $j \sim j_0 $ is experimentally possible. In
this regime the calculated magnetization profile (figure~2) shows a
deviation of $\approx 20^o$ on the boundary. Detection of the effect
is difficult because the spatial resolution of existing probes exceeds
$L_m$.  For a quantitative measurement of the deflection angle on the
boundary, element specific X-ray magnetic circular dichroism (MXCD)
\cite{Gourney} could be used.  A single layer of different magnetic
element can be put on the boundary in the process of the film growth.
Such a layer will not much disturb the overall magnetization profile.
The MXCD signal of the additional layer could be separated from the Mn
signal and thus $\theta(0)$ can be measured.  The
time-dependent approach developed here can be applied to other
experimental geometries, e.g. to find the continuum analogies for the
devices proposed in \cite{slon}.

The authors are greatful to J.Slonczewski for discussion and providing
his paper before publication, to L.P.Pryadko, A.J.Millis, V.N.Smolyaninova
and K.Pettit for discussions.  Research at Stanfrod University was
supported by the NSF grant DMR-9522915.

\end{document}